# Unique Dielectric Behaviour and Anomalies in Nanoconfined Liquids

**Sayantan Mondal**[(1),*] and **Biman Bagchi**[(2),*]

[(1)] *Department of Chemistry and Chemical Biology, Harvard University, 12 Oxford Street, Cambridge, Massachusetts 02138, U.S.A.*

[(2)] *Solid State and Structural Chemistry Unit, Indian Institute of Science, C V Raman Road, Bengaluru, Karnataka 560012, India.*

*Corresponding authors' email: smondal@fas.harvard.edu, bbagchi@iisc.ac.in

## Abstract

The dielectric properties of a bulk dipolar liquid have been subjects of intense interest during the past decades. A surprising result was the discovery of a strong wavenumber dependence in the bulk homogeneous state. Such behaviour seems to suggest the possibility of a strong system size dependence of the dielectric constant (DC) of a nanoconfined liquid, although details have been revealed only recently. Dielectric properties of nanoconfined water indeed show marked sensitivity not only to the size and shape (dielectric boundaries) of confinement but also to the nature of surface-water interactions. For geometries widely studied, namely, water confined in a narrow slit, nanocylinder, and nanospherical cavity, the asymptotic approach to the bulk value of the DC with the increase in confinement size, is found to be surprisingly slow. This seems to imply the appearance of *a dipolar cross-correlation length,* much larger than the molecular length-scale of water. In narrow slit and narrow cylinder, the dielectric function becomes both inhomogeneous and anisotropic, and the longitudinal and transverse components display markedly different system size dependencies. This sensitivity can be traced back to the dependence of the DC on *the ratio* of the mean square dipole moment fluctuation to the volume of the system. The observed sensitivity of collective dipole moment fluctuations to the length scale of confinement points to the possibility of using DC to estimate orientational correlation length scale which has been an elusive quantity. Furthermore, the determination of volume also requires special consideration when the system size is in nanoscale. We discuss these and several other interesting issues along with several applications that have emerged in recent years.

# I. Introduction

Dielectric constant is a response function. It provides crucial information about the polarization created within a system in response to a perturbing external electric field. Dielectric constant also finds wide use in modified Coulomb's law to model the interaction potential between charged species inside a dipolar liquid and also inside proteins. However, most of these established approaches were developed and practiced for macroscopic systems where the size of the system is much larger than the size of a molecule.[1–3] In such cases, the boundary conditions hardly play a role, so long the size of the system is much larger than the size of the molecule comprising the liquid. Also, one need not be concerned with position dependence of dielectric constant, unless under specific conditions. This condition and several others do not hold in nanoconfined liquids.[4] This leads to many interesting, new and unusual properties which have drawn considerable attention in recent years.

In this Perspective Article, we shall review and discuss recent results on the dielectric properties of nanoconfined liquids constrained to different geometries. When the system size is small, in the nanoscale, we lose the advantage of the isotropy whose presence makes things tractable for bulk systems. Also, we lose the advantage of a wavenumber-based description widely used in analytical theories.[5–8] Nevertheless, considerable progress has been made in several different areas. We shall also review theoretical understanding that has emerged from applications of computer simulations, and subsequent theoretical analyses.

The basic perspective that emerges is that the approach to the asymptotic bulk value of the dielectric constant and total dipole moment relaxation are surprisingly slow (with an increase in the size of the system).[9–12] However, single particle properties like translational diffusion and rotational correlation time approach their bulk values, rapidly, at comparatively smaller sizes. This suggests that the intermolecular correlations that determine the collective properties directly, remain perturbed by the surface. Not only are the details of surface-liquid interactions become important, but they influence two particle correlations with the confined liquid. This area requires more theoretical work, like solution of the integral equations which could possibly be implemented with present day computers.

Response functions are determined by the fluctuations in the conjugate properties of the system.[13,14] The dielectric constant is determined the mean square fluctuation of the total dipole moment, as discussed in more detail below. Since dipole moment is a vectorial quantity, understanding of dielectric constant is complicated. In essence, we need quantification of the orientational cross-correlations which are substantial for water in particular. Since dipolar correlations are long ranged, a part of the system size dependence of dielectric constant steps from the latter, at least for water.

Value of the dielectric constant is useful in theoretical analyses and coarse-grained (CG) simulations of electrostatic interaction which is often represented by a reaction-field model.[15] Since we have often to deal with ions in water in various geometries, it is convenient and popular to use an appropriate dielectric constant to mimic the electrostatic interaction potential between the charges. This approach has been successful in bulk water and other macroscopic systems, like the modelling of interactions within protein interiors. The value of the dielectric constant in protein hydration layer and also in DNA grooves has important theoretical consequences.[16]

Unfortunately, the understanding and the evaluation of the dielectric constant of nanoconfined liquids turned out to be more complicated than that for bulk liquids. The most inconvenient aspect is that the values need to be obtained by simulations as analytical theories that exist are all for bulk homogeneous liquids.[7] The dielectric constant of a bulk dipolar liquid is a collective quantity where the fluctuations in total dipole moment involve contributions from two-particle correlations involving all the molecules of the system. In the bulk limit, we can use a language that uses both isotropy and the large size of the system. In the absence of these limiting conditions, analytical solutions of the orientational pair correlations are not available.

For example, when water is confined within a spherical geometry **(Figure 1c)**, the value of the static dielectric constant is found to be markedly different from its bulk value even when the radius is increased to several tens of nm when the spherical system may contain several thousands of water molecules.[9] The convergence to bulk value is found to be slow. This is indeed surprising because the decay correlation length of two-particle spatial and orientational correlations in bulk liquid is short. But this is one of the peculiarities of dielectric constant. In the case of radius dependence of the total static dielectric constant $\varepsilon(R)$ of the liquid confined within spherical confinement of radius R, we can define a macroscopic dielectric correlation length by

$$\varepsilon(R) = 1 + (\varepsilon_{Bulk} - 1)exp[-(\xi/R)^\alpha] \qquad (1)$$

where ξ is the dielectric correlation length (DCL). The exponent $\alpha$ is a parameter that can describe departure from exponential dependence. Computer simulations show the DCL is rather large (much larger than the molecular length scale).[11]

This result gains relevance because of the lack of the existence long-range spatial correlations in bulk water among the isotropic part of the density fluctuations. What one is interested in here is the existence of spatial correlations among orientational degrees of freedom. That is, correlations among orientational density fluctuations. And these appear to be strongly influenced in nanoconfined liquids, to an extent not fully anticipated.

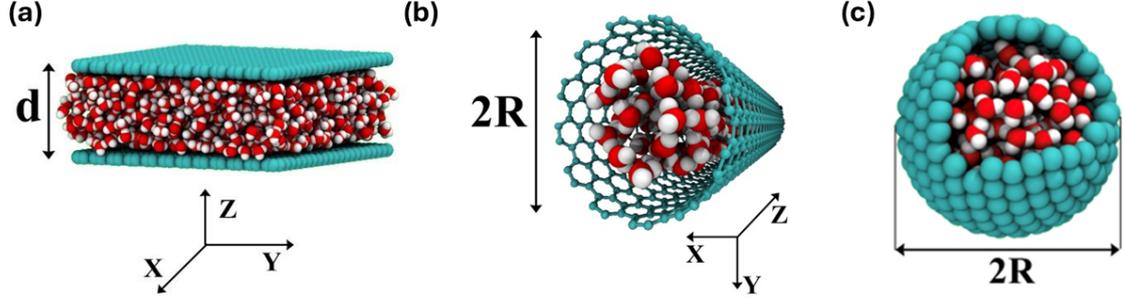

*Figure 1. Water under nano-confinement inside three different geometries: (a) in between two graphene sheets or a nano slit pore of thickness 'd', (b) a nanocylinder of radius 'R' and (c) a nanosphere of radius 'R'.*

When water is confined to a narrow slit or confined inside a tube, the system is inhomogeneous on a mesoscopic length scale. Thus, in a nano slit, one needs dielectric function defined as

$$\boldsymbol{D}(r) = \int d\boldsymbol{r}' \varepsilon(\boldsymbol{r}-\boldsymbol{r}') \boldsymbol{E}(r') \qquad (2)$$

The above equation introduces the dielectric function as a two-point function which is the precise definition obtained from the linear response theory, and expressions have been discussed below. In a bulk liquid, the dielectric function depends only the separation between the two points which in turn allows a smooth definition of wave number dependent dielectric function. These simplicities are lost in an inherently inhomogeneous system we are dealing here.

Nevertheless, we can define a macroscopic dielectric constant because the linear response relation in terms of fluctuation of *the total dipole moment per unit volume*. However, in the constrained small systems that we consider here, the amplitude of fluctuations become much smaller than in the bulk liquid. Eq. (1) is testimony to that effect. Additionally, one observes an inhomogeneous dielectric function where the value depends on the distance from the surface.

In addition to the reduced values, the dielectric function becomes anisotropic in the systems considered. For nano-slits and nano-cylinders, the dielectric function becomes a tensorial quantity with the perpendicular directions (between the plates in nano slit) and the longitudinal direction can have vastly different values.[10,11,17,18]

Several earlier studies were devoted to understand the dielectric response in nanoconfined water, both from theoretical and experimental viewpoints. In one of the pioneering theoretical studies Ballenegger and Hansen investigated the dielectric response of a model polar fluid near a neutral interface by using molecular dynamics simulations.[17] With the help of Kubo's linear response theory, they were able to derive the expressions for spatially resolved dielectric permittivity, $\varepsilon(r)$, for spherical and slab geometries. They found pronounced oscillations in $\varepsilon(r)$ for both the longitudinal and transverse components with the former being ill-defined near the interface due to the emergence of non-local nature [Eq. (2)]. Prior to the above-mentioned study, Senapati and Chandra computed the static dielectric constant of water inside spherical nanocavities. They reported an approximately 50% reduction in $\varepsilon_0$ when the radius of the nanocavity is 6 Å.[19] Unfortunately, the theoretical formalism used suffered from unphysical assumptions (originally by Berendsen[20]) which do not hold in the nanoscopic world.[9]

Later, in a series of studies Netz,[18,21–23] Aluru,[24,25] Marx,[26] Gekle,[27] Maiti,[28] Jalali,[29,30] Zhang,[31] Ghoufi,[32] Matyushov[33] and others further explored the salient aspects of the dielectric response of dipolar liquids (primarily water) under nanoconfinement, with the help of fluctuation theory and molecular simulations. In a pioneering experimental study Geim and colleagues, as a first, measured the out-of-plane dielectric permittivity of water, confined within hexagonal boron-nitride (hBN) slit-pores.[12] They reported that $\varepsilon_\perp$ can be as low as 2 for strong confinements. We have discussed this particular study later in detail [**Section IV.A**].

Let us now discuss this aspect in a bit more detail. If **M(r**,t) is the local dipole moment vector in a small region, then the correlation function is interested is a tensorial quantity given by Eq. (3) as follows

$$C_{MM}(\boldsymbol{r},\boldsymbol{r}',t) = \langle \vec{M}(\boldsymbol{r},0).\vec{M}(\boldsymbol{r}',t)\rangle \qquad (3)$$

In neat liquid water, these correlations decay within a small spatial and short temporal separations. *Thus, the strong effects of the surface on dielectric response are not expected a priori and is hard to explain*. The large dielectric constant of water certainly points to correlated large orientational fluctuations of the dipole moments. One can indeed consider the anomalous slow convergence of static dielectric constant of spherically nanoconfined water to originate from perturbed total moment fluctuations. The latter in turn arises from the alteration in orientational correlations, particularly near the surface.

In a different illustration of the peculiarities presented by nanoconfined liquid systems, the dielectric constant of water nanoconfined in a slit pore (**Figure 1a**) has found to exhibit surprising anomalies.[11] Because of the geometry, the dielectric constant in this system is anisotropic, and is described by two components, transverse and longitudinal. The transverse component describes screening across the narrow slit and is more sensitive to confinement. We shall define these functions with exact expressions below. Another important system, although relatively less explored in the context of dielectrics, is water confined inside a nanocylinder (**Figure 1b**).[10] Here one also obtains two unique components, namely parallel/axial and perpendicular/radial. The latter is more sensitive to the size and volume of the confined geometry.

An important ingredient to understanding finite size effects in correlated condensed phase systems could benefit from understanding the wavenumber ($k$) dependence of response functions, like dielectric constant and structure factors. These are given by space and time-dependent correlation functions. When these correlations do not decay fast with length, we get pronounced finite size effects. Such effects have been well-known in critical phenomena.

Indeed, dielectric constant of a dipolar liquid exhibits strong wavenumber dependence (**Figure 2**).[5] The longitudinal component (the component parallel to the wave number vector) gives a negative value at the intermediate wave number. This might appear to be surprising but is related to the following two factors: (i) the relevant physical quantity is $1 - 1/\varepsilon_L(k)$, which is never negative because $\varepsilon_L(k)$ is never less than unity. (ii) The reason for the negative value of the dielectric constant is due to the emergence of orientation correlations at intermediate wave number (in the region where the static structure factor is peaked in dense liquids).

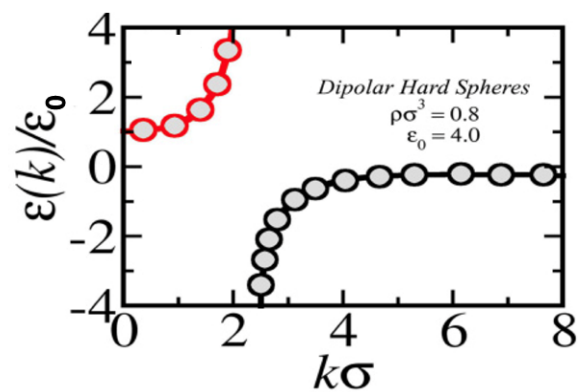

*Figure 2. Wavenumber(k) dependence of the static dielectric function [ε(k)] for dipolar hard spheres at a fixed density. The ε(k) diverges at intermediate values of k and asymptotically approach towards a small negative value [= $1 - 1/(1 - 3Y)$ where $Y = (4/9)\pi\beta\mu^2\rho_0$] at large k. However, divergences do not significantly affect solvation and other properties because the physical quantity is 1- 1/ϵ$_L$(k). The figure is used from Ref [9].*

The question then naturally arises: do confined liquids experience finite size effects through wave number dependence of bulk dielectric constant? That is, finite range of intermolecular correlations influence the dielectric behavior of confined liquids when the length of confinement is of the range of correlation. Alternatively, the surface effects may be unusually long ranged. These two effects can be present simultaneously.

The interference between surface induced alternation of properties and that due to the finite size of the system has been discussed in the context of water confined within cyclodextrin cavity. Experiments find that for large enough size of the cavity, say, for example ten times the diameter of the water molecules, there appears a pool of water at the centre that exhibits dynamic properties of bulk water. There is, however, an issue that has often been ignored. While single particle properties, such as rotational relaxation time constant (measured by fluorescence depolarization or NMR), and translational diffusion coefficient approach bulk values rapidly, *the same seems not to be the case for collective properties*. In particular, dielectric constant appears to be extra sensitive to the size of the system. This difference between single particle property and collective property indicates the role of intermolecular correlations.

In fact, the marked difference between the dynamical characteristics of the single particle rotational motion and those of the many-body collective total dipole moment reveals the existence of long range correlations that seem to play intriguing, not yet fully understood role in the dielectric behaviour of dipolar liquids. As mentioned above, when confined to a spherical geometry, we do have an *exact* expression for the static dielectric constant ($\varepsilon_0$), in the Clausius-Mossotti equation (Eq. 4 and 5).[1]

$$\frac{\varepsilon_0 - 1}{\varepsilon_0 + 2} = \frac{4\pi}{3V}\alpha \tag{4}$$

Here V denotes the volume and $\alpha$ is the static macroscopic polarizability. Now, one utilizes the linear response theory by Kubo,[34] to express $\alpha$ in terms of total dipole moment fluctuation, $\langle \delta M^2 \rangle$. Therefore, in the linear response regime, $\alpha$ becomes $\langle \delta M^2 \rangle/3k_BT$. Hence, the Clausius-Mossotti equation can be re-written as

$$\frac{\varepsilon_0 - 1}{\varepsilon_0 + 2} = \frac{4\pi}{9Vk_BT}\langle \delta M^2 \rangle \tag{5}$$

For a well-sampled ergodic system $\langle M \rangle = 0$, therefore $\langle \delta M^2 \rangle$ and $\langle M^2 \rangle$ can be used interchangeably. In Eq. (5), $\delta M = M - \langle M \rangle$.

The value of the resultant dielectric constant depends delicately on the ratio $\langle M^2 \rangle/V$, where V is the volume experienced by the dipolar liquid. Earlier, we have remarked on the influence of the surface on $\langle M^2 \rangle$. However, in nanoscopic systems, the determination of the volume (V) also poses a serious problem. In small systems, the size of water molecule or the atoms constituting the surface is no longer negligible. This raises interesting issues that needs to be solved. Since there are alternate ways to determine the volume accessible volume, the value of the dielectric constant may differ. This also raises the issue of the utility of translating a function defined in the macroscopic world to the microscopic world. When confined to a narrow slit, once again faces a similar situation when we attempt to determine the dielectric constant in the direction from one plate to the other (that is, the out-of-plane component, $\varepsilon_\perp$), that is, the direction from one plate to the other, denoted by Z-direction.

One often finds that when measured by single particle dynamical properties, the influence of the surface is found to decay fast, within a couple of layers. For example, the rotational correlation time of single water molecules attain the bulk value in the third layer of the water confined within a narrow slit. The same is observed in the protein hydration layer. This is to be contrasted with the slow, long decay of the collective quantities, especially the dielectric function.

In this Perspective article we discuss the fundamental issues that determine the variation of the dielectric constant with the size and shape of the system, by considering three different shapes. In particular, we explore the following question: what could be the reason for the ultralong decay of the surface effect on the collective properties well into the bulk? As mentioned, the same effect has been observed in all the cases studied. In contrast, the single particle properties attain bulk values within 3-4 layers.

## II. Inhomogeneous dielectric function

It was Peter Debye who in his classic monograph on "Polar Molecules"[35] pointed out that dielectric constant of a dipolar liquid can get significantly reduced near an ion and become a function of distance of separation from the ion [**Figure 3**]. The basic idea can be appreciated as follows. If we adopt a CG description, we can describe the collective properties by drawing concentric spheres with an ion at the centre. If one now considers the static dielectric constant in the spherical shells (from the electric polarization created by an external electric field), one would find a lower value in the first shell than the bulk, with values at the concentric spherical regions approaching the bulk value as we move away from the proximity of the ion. The reason is that the water molecules in the first shell undergo less dipole moment fluctuations as they are under the influence of the electric field of the ion. Similar considerations apply to the protein hydration layer. Reduced dielectric constant implies reduced local screening of an electric field between two charged species. As mentioned above, the reduction is a consequence of reduced orientational mobility due to the constraint enforced by the ion or the protein. Interestingly, the opposite scenario of an enhanced dielectric constant has also been observed in nanoconfined water, in nano slits/pores, in the parallel/axial direction,[10,18,22,32] as we discuss below.

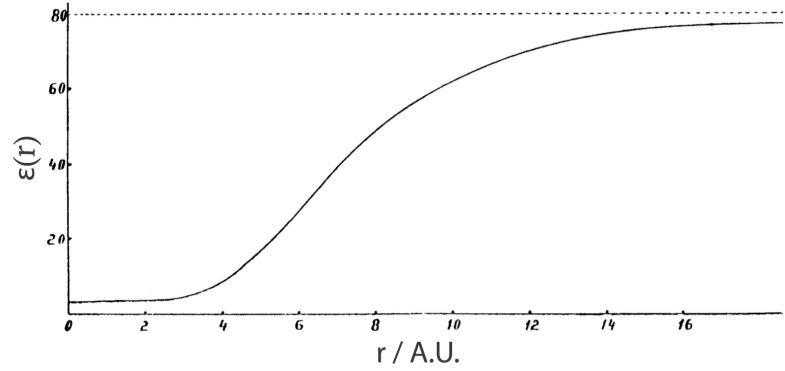

*Figure 3. Position dependent dielectric constant [ε(r)] of water as a function of distance (r) from a monovalent ion. The figure is adapted from the book titled 'Polar Molecules' by Peter Debye, 1929. (Ref [35])*

These results gain relevance in the context of nanoconfined water where multiple new features make their presence felt. First, the dielectric function can become anisotropic in nonspherical systems. And in all systems, in nano scales, the propagation of polarization from opposite surfaces can interfere,[36] leading the confined liquid to exist in a state different from the bulk liquid. Depending on the size of the confinement, the interference between the propagating correlations can be constructive or destructive, leading to more or less reduction of the dielectric constant from the bulk value. Second, the interplay between the surface effects due to the finite size of the system and inherent intermolecular correlations can lead to completely new features which cannot be anticipated *a priori*.

Castner *et al.* adopted a model[6] where the inhomogeneous dielectric constant was decomposed into shells, with each shell having a distinct dielectric constant $\varepsilon_k$, with integer k denoting distance of the shell from the centre. That is, $k = 1$ is the nearest or the first shell. When solved for the reaction field due to the ion and a point dipole in the centre of a sphere, the resultant solvation energy and solvation dynamics was found to be quite different from that predicted by a continuum model in the bulk without such a radially decomposed solvent polarization structure.

# III. Hydrogen bond frustration at the surface: The occurrence of dangling bond

The extended hydrogen bond network of water gets frustrated at a surface, whether the surface is hydrophobic or hydrophilic. A detailed study of such effects on water at a hydrophobic surface has been carried out previously by using the two-dimensional Mercedez-Benz (MB) water model.[37] It was pointed out that *in an attempt to retain the maximum number of hydrogen bond at the surface (or, minimize the frustration), it will point one hydrogen towards the surface (***Figure 4a***), and thus form a dangling hydrogen bond.*[38] In a planar surface, formation of such a dangling hydrogen bond minimizes the loss of hydrogen bonds, that is, the frustration, in the hydrogen bond network. The opposite scenario, where two arms of the MB water are pointed towards the surface (**Figure 4b**), becomes unfavourable. It however introduces an orientational order (**Figure 4c**) which decays as we move away from the surface as discussed earlier, and it seems to play an important role. This orientation order propagates from surface inwards the other confining surface and one finds an interference between the two opposite orders. However, the amplitude and extent of such order is not clear a priori and the effect of this propagation of order and subsequent annihilation of the same, on dielectric properties is non-trivial to compute. One conclusion however can be surmised from the lack of the effect of this order on the single particle properties, like rotation and translation, is the absence of any strong local "caging" potential emanating and propagating from the surface(s).

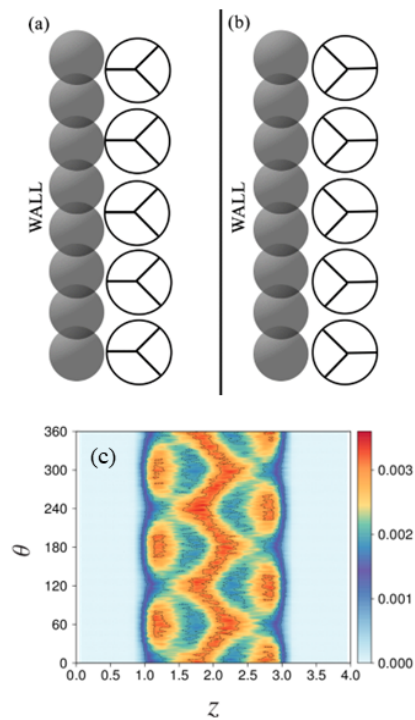

*Figure 4. The relative orientations of two-dimensional Mercedes-Benz (MB) water model near hydrophobic walls/rods. (a) The most preferred orientation where the MB molecules sacrifice one hydrogen bond but manage to retain the formation of two hydrogen bonds with the next layer. (b) An unfavoured orientation where the MB molecules sacrifice two hydrogen bonds and manage to form only one hydrogen bond with the next layer. (c) Orientation map of MB water in the region confined between two hydrophobic rods, at an inter-rod separation distance of 4.0 (reduced units). Here $\theta$ is the angles made by the MB water arms with the normal of the rods. Panel (c) is adapted from Ref [38].*

The observed reduction in the value of the dielectric constant in nanoconfined water certainly arises from the quench of fluctuation in the total dipole moment. The probability distribution of the magnitude, P(M), of the total dipole moment is largely Gaussian and the width of the distribution, the standard deviation, essentially gives the dielectric constant. As one can easily surmise from an expression of DC in terms of the standard deviation or mean square fluctuation, the DC is sensitive to the value of the standard deviation. Even a small error in the mean square dipole moment fluctuation can lead to a large error in the value of the static DC.

It was discussed earlier that the reduction in the mean square fluctuation of the total dipole moment could arise from cancellation of the dipole moment contributions from the opposite average orientations of the individual molecular dipoles near the two opposite ends.[9,11,36,38] This is due to presence of dangling bonds which are oriented in the opposite direction, naturally. In Figure 4(c), a preferred orientation can be observed adjacent to the walls which is a result of Figure 4(a) like orientation. Such orientational bias at the interface propagates inward through the subsequent water layers towards the middle of the confinement. This effect creates orientationally frustrated water molecules in the middle. One can further conjecture that the surface gives rise to an orientational caging potential that acts on the position dependent locally evaluated collective moment, with caging opposite at the two ends.

There is another important aspect that has not been addressed adequately is the issue of the total volume (V) that is experienced by the confined liquid. In the macroscopic system, volume is well-defined. However, in the nanoworld, this volume is somewhat ambiguous. At small scales, volume accessible to a molecule is determined by the intermolecular interactions. For example, a water molecule confined within a sphere made of equally spaced small spheres on a curved surface experiences a variation of potential on sub-nm scale. One might recourse to the traditional method of using the radial distribution [Eq. (6)] for an estimation of the volume accessible to the water molecule.

$$\frac{1}{N}\int_0^{R_{eff}} dr\, 4\pi r^2 \rho(r) = 1 \tag{6}$$

Here, $\rho(r)$ is the radial density profile of the liquid, N is the total number of water molecules, and $R_{eff}$ is the effective radius of the system which is less than the actual geometric radius (R). Hence, for a nanosphere the effective volume becomes $V_{eff} = (4/3)\pi R_{eff}^3$. For example, in spherical nanocavities, $R_{eff}$ was found to be ~0.7 Å less than the actual R.[9] Likewise, in nanocylinders the $R_{eff}$ was found to be ~1.8 Å less that the actual R.[10] In slit pores the effective distance between plates, $d_{eff}$ was found to be ~1.4 Å less that the geometric distance.[11] This is not unique to the determination of dielectric constant. We shall return to this point again later.

## IV. Theoretical description of the spatial heterogeneity

The dielectric permittivity becomes spatially heterogeneous under nanoconfinement, we need to define a spatially resolved dielectric profile, $\varepsilon(\mathbf{r})$, and a rigorous statistical mechanical definition becomes nontrivial. In this section, we follow the theoretical frameworks developed by Hansen et al.[17] and later by Netz et al.,[18,21] to briefly discuss the formulation of $\varepsilon(\mathbf{r})$ in nanoconfined polar liquids.

If $\Delta \mathbf{E}(\mathbf{r}) = \mathbf{E}(\mathbf{r}) - \mathbf{E}_0(\mathbf{r})$, is the difference between the local electric fields inside the cavity in the presence and in the absence of the external field, the induced polarization density $[\Delta \mathbf{P}(\mathbf{r})]$ can be expressed as

$$\Delta \mathbf{P}(\mathbf{r}) = \frac{1}{4\pi} \int_{D_{Cavity}} d\mathbf{r}' \underline{\chi}(\mathbf{r},\mathbf{r}') \Delta \mathbf{E}(\mathbf{r}') \qquad (7)$$

where $\underline{\chi}(\mathbf{r},\mathbf{r}')$ is the non-local dielectric susceptibility tensor. Eq. (7), under the slow modulation limit, reduces to

$$\Delta \mathbf{P}(\mathbf{r}) = \frac{1}{4\pi} \underline{\chi}(\mathbf{r}) . \Delta \mathbf{E}(\mathbf{r}) \qquad (8)$$

where $\underline{\chi}(\mathbf{r})$ is the local dielectric susceptibility tensor, which is in turn related to the dielectric permittivity tensor as $\underline{\chi}(\mathbf{r}) = \underline{\varepsilon}(\mathbf{r}) - \underline{I}$. According to Kubo's linear response theory, $\Delta \mathbf{P}(\mathbf{r})$ generated by an external field ($E'$) and the cavity field $E^c$ are related by the following relation

$$\Delta P_\alpha(\mathbf{r}) = \beta \sum_{\gamma=x,y,z} [\langle m_\alpha(\mathbf{r}) M_\gamma \rangle - \langle m_\alpha(\mathbf{r}) \rangle \langle M_\gamma \rangle ] E^c_\gamma \qquad (9)$$

where $\beta$ is $(k_B T)^{-1}$, $\boldsymbol{m}(\boldsymbol{r})$ is the radial component of the microscopic polarization density given by $\boldsymbol{m}(\boldsymbol{r}) = \sum_{i=1}^{N} \boldsymbol{\mu}_i \delta(\boldsymbol{r} - \boldsymbol{r}_i)$, and $\boldsymbol{M}$ is the total dipole moment given by $M = \int d\boldsymbol{r}\, \boldsymbol{m}(\boldsymbol{r})$. Eq. (9) is a generalization of Kubo's LRT for anisotropic systems where it takes a more involved form. However, Eqs. (7) to (9) are not sufficient to derive the fluctuation formulae for the dielectric constant in nanoconfinement, as they do not contain information on the geometry or the dielectric boundaries.

Now, for a slab geometry, one usually considers the Z-direction to be perpendicular to the walls. Maxwell's equations suggest that $\nabla \times \boldsymbol{E}(z) = 0$ and $\nabla \cdot \boldsymbol{D}(z) = 0$, where **E** and **D** are related by $\boldsymbol{D} = \boldsymbol{E} + 4\pi \boldsymbol{P}$.[1,2] Therefore, $E_\parallel^c = E_\parallel'$ and $E_\perp^c = \varepsilon_\perp E_\perp'$. Hence, the following fluctuation formulae [Eq. (10) and (11)] can be derived

$$\varepsilon_\parallel(z) = 1 + 2\pi\beta[\langle \boldsymbol{m}_\parallel(z) \cdot \boldsymbol{M}_\parallel \rangle - \langle \boldsymbol{m}_\parallel(z) \rangle \cdot \langle \boldsymbol{M}_\parallel \rangle] \tag{10}$$

$$1 - \frac{1}{\varepsilon_\perp(z)} = 4\pi\beta[\langle \boldsymbol{m}_\perp(z) \cdot \boldsymbol{M}_\perp \rangle - \langle \boldsymbol{m}_\perp(z) \rangle \cdot \langle \boldsymbol{M}_\perp \rangle] \tag{11}$$

In the above equations, the contribution of inter-grid cross-correlations was also taken into consideration. These equations can be used to numerically obtain the grid-wise dielectric constant inside a slit-pore.[17,21]

In a spherical geometry, one assumes a radially emanating electric field $[\boldsymbol{E}'(\boldsymbol{r}) = (q/r^2)\hat{\boldsymbol{r}}]$ from the centre of the sphere. Hence, the dielectric permittivity profile becomes

$$1 - \frac{1}{\varepsilon(r)} = 4\pi\beta \int_{D_{cavity}} d\boldsymbol{r}'\, \langle m(\boldsymbol{r}) m(\boldsymbol{r}') \rangle \left(\frac{r}{r'}\right)^2 \tag{12}$$

Likewise, for a nanotube one assumes a radially emanating field from a dimensionless wire placed along the axis of the nanotube with a charge density of $q/L$ (L is the length of the nanotube). The fluctuation relations for the axial $[\varepsilon_z]$ and radial $[\varepsilon_r]$ components are

$$\varepsilon_z(r) = 1 + 2\pi\beta L \left( \langle m_z(r) \int_0^\infty dr' r' m_z(r') \rangle - \langle m_z(r) \rangle \langle \int_0^\infty dr' r' m_z(r') \rangle \right) \quad (13)$$

$$1 - \frac{1}{\varepsilon_r(r)} = 2\pi r L \beta \left( \langle m_r(r) \int_0^\infty dr' m_r(r') \rangle - \langle m_r(r) \rangle \langle \int_0^\infty dr' m_r(r') \rangle \right) \quad (14)$$

Although theoretically intriguing, the spatially resolved dielectric permittivity profiles, $\varepsilon(r)$ is hard to observe experimentally. Experiments can only capture the effective dielectric constant, $\varepsilon_{eff}$. However, the knowledge of $\varepsilon(r)$ could be useful in the modelling of electrostatic interactions in CG simulations under confinement or near interfaces.

## V. Dielectric behaviour of water confined within nano-containers of different geometry

In the following we discuss dielectric constant of confined liquids, particularly water, in three geometries. We discuss the available experimental results wherever possible. Much of the quantitative understanding has been acquired through computer simulations.

### A. Between narrow slit-pores

This system has drawn considerable attention in modern times because of its importance in both materials science (batteries) and biology (plants). In the standard model of water within narrow slits, a few layers of water are confined between two parallel plates which extend to macroscopic scales. Thus, this system is reminiscent of a (quasi) two-dimensional system. This model has been a subject of studies over many decades. The earliest example of experimental studies of this system is the study of Israelchevelli who tried to understand the amplitude and extent of hydrophobic attraction among two paraffin-coated parallel plates.[39] This experiment reveals the existence of a long-range attractive force between the surfaces which has been termed hydrophobic attraction. An elegant theory has been developed to explain this attractive force.[40] However, the initial experiments focussed on distance of the order of 10 nm or more. More recently experiments have been carried out with much narrower slits, with emphasis on measuring the value of the static dielectric constant across the plates that we call the Z-axis.

In an interesting experimental study, Geim *et al*. measured the out-of-plane (or the perpendicular) static dielectric constant ($\varepsilon_\perp$) of water sandwiched between hBN sheets.[12] For lower values of the inter-slab separation (d < 5 nm), they obtained surprisingly low values of $\varepsilon_\perp$ (even as low as 2). In order to recover the bulk value of 80, they needed the inter-slab separation to increase beyond 100 nm. That is, the value of the measured dielectric constant was much below 80 even at a separation of 20 nm or so. They attributed these observations to the presence of electrically dead layers (EDL), which could be 2-3 molecular diameters thick. These EDLs possess vanishingly small orientational polarizability.

Theoretical studies have addressed this unusual result. The dielectric tensor for slab confinement (or slit-pore) is anisotropic with two unique eigenvalues, namely, the parallel ($\varepsilon_\parallel$) and perpendicular ($\varepsilon_\perp$) components. One can probe the latter by placing an external field perpendicular to the slab surfaces. The cavity field becomes, $E_{Cavity} = E_{Ext}/\varepsilon_\perp$. Therefore, $\varepsilon_\perp$ can be defined as

$$1 - \frac{1}{\varepsilon_\perp} = \frac{4\pi}{Vk_BT}\langle\delta M_\perp^2\rangle \qquad (15)$$

where $M_\perp$ is the perpendicular (with respect to the slab plane, Z-direction by convention) component of the total dipole moment. The parallel component can be defined as

$$\varepsilon_\parallel = 1 + \frac{4\pi}{Vk_BT}\langle\delta M_\parallel^2\rangle \qquad (16)$$

where $M_\parallel{}^2 = \frac{1}{2}(M_X{}^2 + M_Y{}^2)$. The detailed derivation of Eq. (15) and (16) can be found elsewhere.[11]

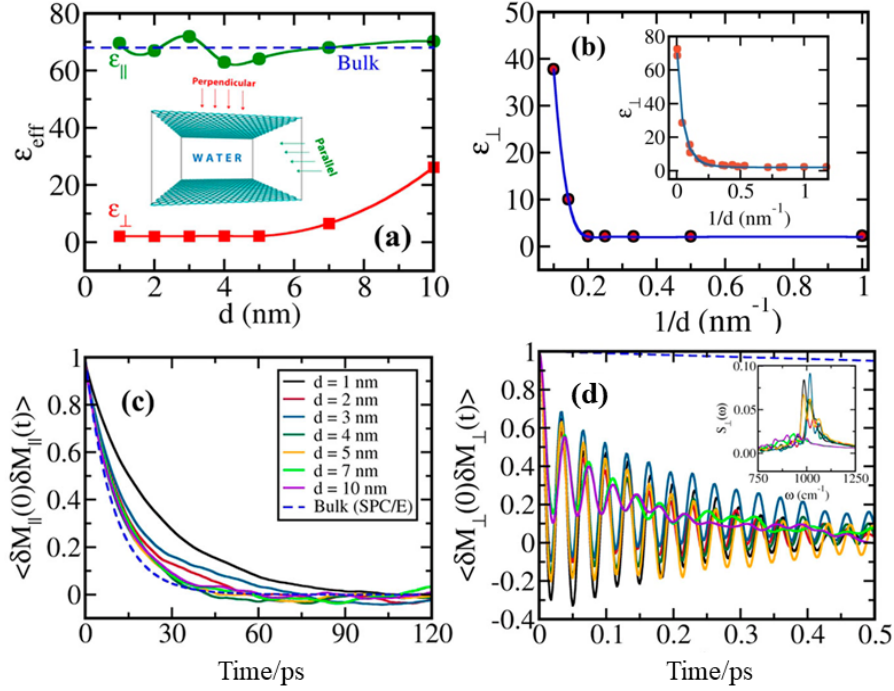

Figure 5. (a) The parallel and perpendicular components of the effective macroscopic SDC of water under slab-confinement (as shown in the inset) with an increase in inter-plate distance (d). The parallel component converges rapidly to the bulk value (~68 for SPC/E) after the initial oscillatory behaviour, whereas the perpendicular component exhibits an extremely slow convergence with an increase in d. A similar slow convergence of $\varepsilon_\perp$ was captured in dielectric microscopy experiments by Geim et al. (b) The perpendicular component of the static dielectric constant ($\varepsilon_\perp$) for slab geometries against the inverse of the inter-slab separations. The correlation length scale is obtained by fitting the $\varepsilon_\perp$ vs 1/d plots obtained from simulations and experiment (inset) with a stretched exponential function of the form Eq. (1). The plots show a similar initial slow convergence followed by a steep increase. The obtained correlation lengths ($\xi$) are approximately 9 nm (simulations) and 22 nm (experiments). Relaxation of the (c) parallel and (d) perpendicular components of the total dipole moment time correlation function of slab-confined water. Dielectric relaxation in the parallel direction is slower than the bulk (blue dashed line) and converges to the bulk decay pattern with an increase in inter-plate separation, d. On the contrary, the perpendicular dielectric relaxation exhibits ultrafast and oscillatory relaxation that differs by two orders of magnitude compared to the bulk. Interestingly, all the oscillatory time correlations correspond to a frequency of ~1000 $cm^{-1}$ [inset, (d)]. As we increase d to 10.0 nm, the oscillation reduces in amplitude, but the decay remains ultrafast. Figures on all the panels are adapted from Ref[11].

Note the difference between the above two expressions. In almost all the applications of the narrow slit, only the perpendicular component is relevant because experimentally we place the field or the charges across the confining plates. *From the molecular dynamics perspective, one needs to be cautious regarding the use of electrostatics. Instead of the usual three-dimensional Ewald summation for the long-range electrostatics, one needs to consider the modified/corrected Ewald sum (Ewald-3dc) with the recommended system specifications.*[41]

We calculated the two components of the static dielectric constant (SDC) of water confined in narrow slit pores for different inter-slab distances, $d$ [**Figure 5a**]. For the parallel component ($\varepsilon_\parallel$), we observed a slight enhancement in the SDC compared to the bulk value (~68 for SPC/E water at 300 K). On the other hand, we observed a slow convergence of the SDC with increasing $d$ towards the bulk value along the direction of the confinement ($\varepsilon_\perp$). This is in agreement with experiments. However, in computer simulations, we could not go to the value required to obtain the asymptotic bulk value. Although the change in '$d$' casts a profound effect on $\varepsilon_\perp$, the change in the twist-angle between two graphene sheets (that is, different Moire patterns) has weak or no effect on $\varepsilon_\perp$.[42] Interestingly, the change in twist angle can affect the material properties of graphene, but could not affect the dielectric properties of water sandwiched in between. In an earlier study by Netz and colleagues, a giant dielectric response ($\varepsilon_\parallel \sim 400$) near the interface was observed in the parallel direction for water confined between dipalmitoylphosphatidylcholine (DMPC) lipids.[22] However, such a giant response was not observed for decanol or digalactosyldiacylglycerol (DGDG) surfaces. This result indicate the effect of surface polarity on the dielectric response.

In order to quantify the correlation length scale, we plot the calculated values of $\varepsilon_\perp$ against the inverse of the inter-slab separation distance in **Figure 5b**. We fit the data by using a stretched exponential function as in Eq. (1) to extract the correlation length ($\xi$). The correlation length is found to be approximately equal to 9 nm with α=3.5. The values of the SDC are extremely low for smaller slits. It becomes as low as 2 for smaller values of the inter-slab separation. This region matches well with the experiments of Geim and co-workers.[12] However, the correlation length obtained from the same experiments is approximately 2.5 times larger, that is, $\xi \sim 22$ nm (**Figure 5b**, *inset*) than that obtained from simulations. The origin of this incongruity may lie in the choice of force field.

Interestingly, the time correlation function of the perpendicular component of dipole moment fluctuation exhibits ultrafast decay with pronounced oscillations [**Figure 5d**]. However, the same for the parallel component exhibits slower relaxation compared to the bulk and converges to the bulk pattern with an increase in the size of the confinement [**Figure 5c**].

The amplitude of the observed oscillations decreases with an increase in *d*, but the timescale of relaxation remains unchanged. We discovered the ubiquitous presence of a frequency of ~1000 cm$^{-1}$ from discrete Fourier transformations [**Figure 5d**, *inset*]. The frequency is omnipresent for systems with d ≤ 5 nm and gradually diminishes for systems with large inter-slab separations. We attributed its origin to the emergence of an effective localization/caging potential in the perpendicular direction.

Oscillations in a collective mode arise from underdamped motions. There is evidence of the presence of dipolaron modes (long-lived collective excitations)[43,44] that could lead to the oscillatory nature of $\langle \delta M_\perp(0) \delta M_\perp(t) \rangle$. As the perpendicular component does not possess the k=0 mode, the full dipolaronic character cannot be present. However, we can still observe the oscillations as the collective modes along the perpendicular direction is coupled to the high-frequency modes of the liquid which experiences less frictional resistance. Chandra and Bagchi studied the dipolaron modes in dense dipolar liquids.[43] Such modes were also observed in the intermediate wavenumber regime. It is difficult to observe the dipolaron modes in normal liquids due to the fact that motions are overdamped and the excitations are short lived. Nevertheless, the presence of a strong confinement can favour their existence. Earlier, such pronounced oscillations in $\langle \delta M_\perp(0) \delta M_\perp(t) \rangle$ were also connected to the dominance of librational modes.[23]

In the beginning of this perspective, we discussed the case of dipolar hard spheres that exhibit a negative value of the dielectric constant at intermediate wavenumbers [Figure 2]. In the context of nanoconfined water, a few recent studies reported a negative value of the dielectric constant.[28,30,45] Sugahara *et al.* studied water inside ion intercalated MXene nanoslits and found that dipolar polarization of the strongly confined water can resonantly *overscreen* an external electric field to enhances capacitance with a characteristically negative dielectric constant.[45] MD simulations studies by Jalali *et al.*[30] and later by Majumder et al.[28] further explored the aspect of a negative dielectric constant. The former found an anomalous increase in the capacitance with the decrease of radius of the intercalated cations, that they connected to a negative dielectric permittivity.[30] The latter studied monolayer water in graphene slit pores and found the signature of negative out-of-plane dielectric permittivity around the centre of the pore.[28] They attributed this observation to the ordering of water dipoles to form a square ice structure under strong confinement.

A simple theoretical explanation of the observed anomalously slow approach to the bulk value of SDC has been put forward. The nano slab system can be compared to a parallel plate capacitor filled with a dielectric material of varying dielectric constant. For such a system, the capacitance ($C$) is given by $C = \varepsilon A/d$, where 'A' and 'd' are the area of the plates and interpolate distance, respectively. $\varepsilon$ is the dielectric constant of the intervening medium. Now, we consider (as in the previous discussions) that the water between the two plates consists of several dielectric layers, each with distinct dielectric constant. Therefore, the water layer between the slabs can be considered as a combination of parallel plate capacitors in series connection. It then follows that the effective SDC is the harmonic mean of the grid-wise SDCs [Eq. (17)].[46]

$$\frac{1}{\varepsilon_{eff}} = \frac{1}{N} \sum_{i=1}^{N} \frac{1}{\varepsilon_i} \qquad (17)$$

Hence, the electrically dead water layers with vanishingly low SDC make a disproportionately large contribution to the observed value. This provides a semi-quantitatively valid explanation of the observed results. However, a detailed calculation is still missing. It was further shown that the surface layers, although electrically dead, are dynamically quite alive.[46]

In nano slits $\varepsilon_\parallel(z)$ is enhanced and $\varepsilon_\perp(z)$ is diminished at the centre, with the former being negative over an extended range. With the aim of combining these unique features into a description suitable for CG modelling, Netz and colleagues developed a dielectric box model by using effective medium theories.[47] They found that the value of effective dielectric permittivity $\varepsilon_{\parallel/\perp}^{eff}$ is sensitive to the value of the effective length of the box ($L_{\parallel/\perp}^{eff}$), which is similar to what we have discussed in Section III. In narrow slits, $L_{\parallel}^{eff}$ increases compared to the water layer thickness ($L_w$) which indicates an increased efficiency of water as a parallel dielectric. On the other hand, at similar length scale $L_{\perp}^{eff}$ decreases compared to $L_w$ which indicates a decreased efficiency of water as a perpendicular dielectric.

B. **Water inside nanocylinders**

In a cylindrical geometry, the dielectric response also becomes anisotropic with two distinct components- axial ($\varepsilon_z$) and perpendicular ($\varepsilon_{xy}$). The former is straightforward to obtain, as the cavity field is $\varepsilon_z E_{ext}$. Therefore,

$$\varepsilon_z = 1 + \frac{4\pi}{V k_B T} < \delta M_z^2 > \qquad (18)$$

However, for an external field sent perpendicular to the nanotube axis, the cavity field becomes, $E_{Cavity} = [3/(\varepsilon_{xy} + 1)] E_{ext}$. Hence, the perpendicular component can be written as,

$$\frac{\varepsilon_{xy} - 1}{\varepsilon_{xy} + 1} = \frac{4\pi}{V k_B T} < \delta M_{xy}^2 > \qquad (19)$$

where $M_{xy}^2 = (1/2)[M_x^2 + M_y^2]$.

In **Figure 6a**, we plot $\varepsilon_z$ and $\varepsilon_{xy}$ against the inverse of the number of nanoconfined water molecules ($N_{wat}$). For narrow cylinders (that is high $1/N_{wat}$) the axial component gets enhanced from the bulk value. However, the perpendicular component initially exhibits a slow convergence with the increasing size of the nanocylinder followed by a steep increase. In **Figure**

**6b**, $\varepsilon_{xy}$ was plotted against 1/R (R is the radius of the nanocylinder) to extract the correlation length ($\xi$) according to Eq. (1). The value of $\xi$ for cylindrical geometry is found to be approximately 3 nm. The total dipole moment correlation function in the perpendicular direction [**Figure 6c**] was found to be ultrafast (almost 30 times faster relaxation than the bulk). The origin of such anomalous fast relaxation can be understood by evaluating the Kirkwood g-factor, described later.

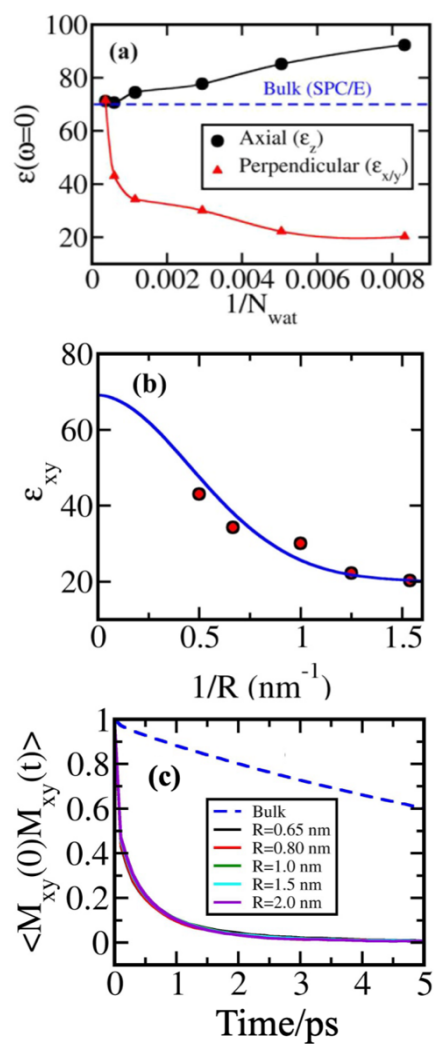

*Figure 6. (a) Two components ($\varepsilon_z$ and $\varepsilon_{xy}$) of the static dielectric constant of water confined inside a nanotube/cylinder, plotted against the inverse of the number of confined water molecules, $N_{wat}$. [adapted from Ref[10].] (b) The perpendicular component of the static dielectric constant ($\varepsilon_{xy}$) for nanotubes of different sizes against the inverse of the radius. The correlation length scale is calculated by fitting the $\varepsilon_{xy}$ vs 1/R plots obtained from simulations with the expression given in Eq. (1). $\varepsilon_{xy}$ shows an initial slow convergence, followed by a steep increase. However, the steepness is less than that of the slab geometry. Here the obtained correlation lengths ($\xi$) is approximately 3 nm with $\alpha$=1.9. (c) Total dipole moment autocorrelation functions along the non-periodic/perpendicular directions for nano-cylinders of different sizes. The relaxation is ultrafast and weakly dependent on the nanotube diameter.*

## C. Spherical Confinement

This is an interesting case and is a case of practical interest in chemistry where systems such as reverse micelles or liposomes can be modelled as water in an enclosure. In a spherical confinement, water experiences multiple influences that perturb its natural bulk structure and dynamics. It is found by accurate simulations, that the SDC of water confined in a nanocavity exhibits a strong dependence on the size with a remarkably low value for nanospheres with smaller radius.

The Clausius-Mossotti equation [Eq. (5)] provides an exact expression for dielectric constant of a liquid confined within a spherical encampment. However, from a molecular simulation perspective, one needs to be careful regarding a couple of important considerations. For example, (i) the nanospherical system should be isolated without the presence of any periodic boundaries and suspended in vaccum, (ii) because of the absence of periodic boundaries one needs to use real space electrostatic calculations for the whole system.

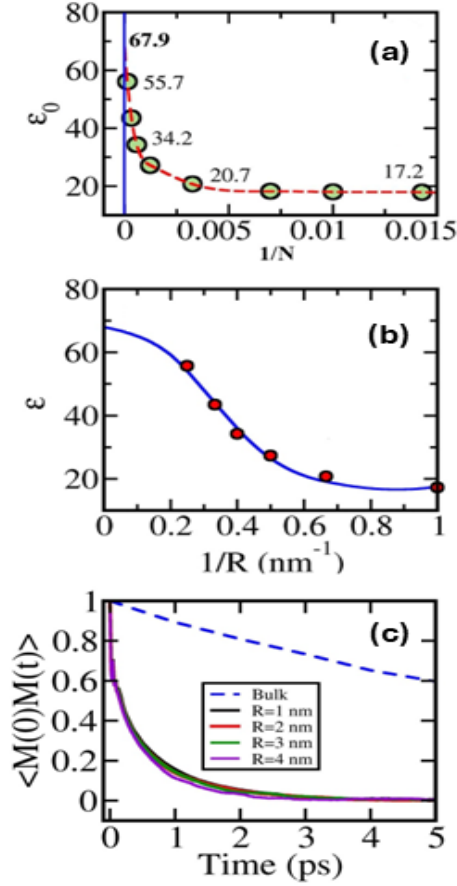

*Figure 7. (a) The static dielectric constant ($\varepsilon_0$) of water under spherical nanoconfinement against the inverse of the number of water molecules (N). The extrapolation shows that $\varepsilon_0$ approaches the bulk value of SPC/E water at 300K (~68) as N approaches ∞ (thermodynamic limit). (b) The unique component of the static dielectric constant ($\varepsilon$) for aqueous nanocavities against the inverse of the radius. The correlation length scale is obtained by fitting the $\varepsilon$ vs 1/R plot from simulations with a stretched exponential function of the same form given in Eq. (1). The plot shows an initial slow convergence, followed by an increase. The correlation length ($\xi$) is approximately 5 nm. (c) Total dipole moment autocorrelation function nano-spherical systems of different sizes. The relaxation is ultrafast and substantially faster than the bulk relaxation (blue dashed line). The relaxation timescales remain almost the same irrespective of the size of the cavity. [Figures of panel (a) and (c) are adapted from Ref [9] and panel (b) is adapted from Ref[11].]*

We plot the SDC against the inverse of the radius (1/R) for the nano-spherical systems [**Figure 7b**] to obtain the correlation length by fitting this to Eq. (1). We obtain $\xi$ to be approximately 5 nm (with $\alpha$=2.5). The obtained value of $\xi$ is quite large compared to the molecular diameter of water (~0.3 nm) and can be considered long-ranged. We note that in spherically confined Stockmayer fluid (SF), $\xi$~2.0 nm which is less than that in water. We attributed the reduction in the correlation length to the absence of long-range electrostatic interactions and hydrogen bond network in SF.

The total dipole moment autocorrelations of nanoconfined water along were found to be ultrafast [**Figure 7c**]. Surprisingly, in all the systems, the relaxation timescales remain almost the same (~20 times faster than the bulk) for systems with different sizes. The origin of this universal anomalous behaviour lies in the low values of Kirkwood g-factor of these systems. It is noteworthy to mention that, for the spherical geometry, an unconstrained extrapolation of $\varepsilon_0$ against the inverse of the number of water molecules (N) produces a value close to 68 at 300 K in the thermodynamic limit (1/N → 0). This value is a more accurate measure of the bulk SDC than from the systems with artefacts arising due to periodic boundaries.

In fact, an earlier study of the dielectric constant of a cubic lattice with point dipoles embedded at lattice sites (known as Lax-Zwanzig lattice) observed a similar slow convergence to the bulk value.[48–50]

## D. Origin of quenched polarisation fluctuation and ultrafast collective relaxation

The orientationally rigid water molecules that reside at the interfacial layer are oriented in the opposite directions, because of the presence of the dangling hydrogen bond as discussed above. This configurator in prone to give rise to two inwardly propagating orientations from the two confining walls and these in turn can give rise to destructive interference among propagating orientations. This destructive interference can create orientationally frustrated water molecules at the centre of the confinement, as discussed extensively for aqueous reverse micelle systems. Such long range interactions can influence the mean square fluctuations in the total (collective) dipole moment.

The Kirkwood g-factor, defined by Eq. (20), serves as a measure of inter-molecular dipolar correlations.[51,52] Here $\langle M^2 \rangle$ is the mean squared total dipole moment of the system, N is the total number of particles, and µ is the single-particle dipole moment (2.35 D for SPC/E water). $g_k = 1$ indicates uncorrelated single-particle orientation, $g_k > 1$ denotes a constructive single-particle dipolar orientation, and $g_k < 1$ denotes single-particle orientations that lead to destructive interference.

$$g_k = \frac{\langle M^2 \rangle}{N\mu^2} \quad (20)$$

The calculated $g_k$ along the non-periodic directions of the systems revealed a substantially attenuated value (an order of magnitude less) compared to the bulk.[9–11] This indicates a destructive inter-molecular interference and can again be attributed to the surface effects. The low values of $g_k$ are intimately connected to the timescales of total dipole moment relaxation. The ratio of the $g_k$ under confinement to the $g_k$ of the bulk is approximately equal to the ratio of the collective orientational relaxation timescales in the two systems.[51]

## VI. Biophysical systems

Water is involved in a large number of biophysical phenomena, not just in protein and DNA solvation and stabilization, but also in the functioning of lipids, and also involved in cellular activities in cytoplasm. The polar response of water layers is important in many systems. In the following, we review a few of the more recent studies.

### A. Dielectric constant of protein hydration layer

The altered polar character of protein hydration layer (PHL) was discovered and discussed by Pethig,[53] Grant,[54] Mashimo[55] and others in later 1970s and early 1980s through both dielectric relaxation and NMR spectroscopies. Frequency dependence of dielectric spectrum revealed the existence of a time scale of the order of 40-50 ps which was much larger than the time of 8.3 ps which is commonly observed in neat water. The new (40-50 ps) time scale much shorter than protein rotation which would be in the time scale of 10 ns or so.

Both the hydrophobic and hydrophilic groups on the surface of the protein interact to modify the extensive hydrogen bond network.[56,57] It is natural that water structure and dynamics become altered. The question to be answered pertains to the extent of distortion from bulk water structure and dynamics. While there is yet no clear consensus, it is generally believed and/or assumed that water surrounding protein gets perturbed due to the influence of the protein up to a few nm from the surface of the solvated protein.[16,58] And the dielectric constant of the water layers adjacent to protein is much smaller than that in the bulk.

However, this field has seen many discussions, and some controversy in recent decades.[59] It started with an influential paper by Wuthrich who presented an approximate measure of the time scale of rotation of water molecules in the hydration layer.[60] His estimate of relaxation time was close to 300 ps which kind of agreed semi-quantitatively with dielectric relaxation estimates, although larger. Much of the controversy is centred around the time scales of rotational motion of water molecules, and of solvation dynamics.[61,62] Experimental studies often characterize the polarity of the hydration layer by a quantitative measure, called Z-score which is measured by the shift in the absorption or fluoresce spectrum of a dye located in the hydration layer. We unravelled the origin of the controversy regarding the relaxation timescales of water in the biomolecular hydration layer by showing the existence of a log-normal distribution in the single particle rotational and translational timescales.[16,59]

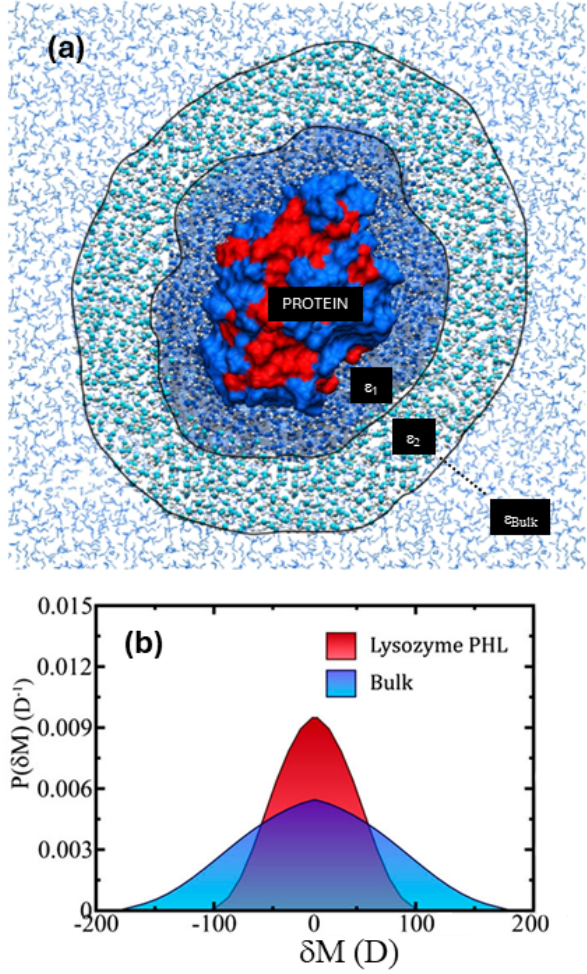

*Figure 8. (a) Cross-sectional view of a protein (lysozyme) in water (SPC/E). The water around the protein is divided into different shells of 1 nm width. The first layer of water molecules is the protein hydration layer (PHL) which is found to exhibit a significantly low effective static dielectric constant (~45). As one moves away from the protein, the static dielectric constant approaches the bulk value (~68 for SPC/E water). (b) The distribution of the total dipole moment fluctuation of the Lysozyme PHL and the bulk SPC/E water at 300K. The distribution for the PHL becomes narrower than that of the bulk which indicates quenched total dipole moment fluctuation of the PHL. Both the figures are taken from Ref [16].*

A few earlier studies used atomistic molecular dynamics simulations of different aqueous proteins to the static dielectric constant (SDC) of the PHL.[16,63] In this study, we divided water around a protein into different shells of 1 nm thickness (**Figure 8a**) and defined an effective local static dielectric constant ($\varepsilon_i^{eff}$, where $i$ is the shell index) in the following fashion.

$$\varepsilon_i^{eff} = 1 + \frac{4\pi}{3k_B T V_i^w} \langle [M_i^w - \langle M_i^w \rangle]^2 \rangle \quad (21)$$

Here we defined a shell-wise SDC, in a heuristic way, when the cross correlation between the total dipole moments of adjacent shells is negligible. The extent of cross-correlation can be estimated by the calculation of Pearson's correlation coefficient which can vary between 0 (0%) and 1 (100%).[64] In our calculations, the extent of dipole moment cross-correlation between adjacent shells generally remained below 10% and hence may be neglected.[16] We found that for different proteins, the SDC of the first shell (PHL) was close to 45 and that of the second shell was close to 55, whereas the bulk value for SPC/E water model is approximately 68 at ambient conditions. As observed earlier, the marked reduction is connected to the quenched total dipole moment fluctuations in the PHL, as depicted in **Figure 8b**. Because of the low SDC and hence less electrostatic screening, the PHL rather facilitates the interaction with substrates or other proteins, which is a prerequisite for protein functions such as enzyme kinetics or even protein aggregation.

### B. Dielectric relaxation in DNA solution

Dielectric relaxation of aqueous DNA solutions was first studied extensively by Oosawa.[65] It has also been investigated, albeit indirectly, by Berg, Sen and coworkers in their experimental studies on solvation dynamics.[66–69] Both dielectric relaxation and solvation dynamics are highly non-exponential.

We include this system here because the observed anomalies have been attributed to the water molecules, and to the counter ions that are confined to the neighbourhood of the DNA. The water that is confined to the minor grooves of DNA have been found to have different orientational order than those in the bulk.[70] The positively charged counterions are found to migrate between the negatively charged phosphate groups and contribute to the polarization relaxation.[71] One can categorize the following sources of non-exponential dielectric relaxation in DNA solutions.

(i) Movement of ions along the DNA chain

(ii) Coupling between water and ions along the DNA chain

(iii) Slow large-scale motions like bubble formation

(iv) Coupling between large scale DNA fluctuations and water and ions.

Oosawa pointed out that the fluctuations in the location of the counterions and their occupancy around the DNA backbone can result in an increase in the observed value of the dielectric constant and can give rise to a markedly slow dynamics. The negatively charged DNA backbone acts as equally spaced traps for the positively charged counterions in the solution. The counterions can add to the polarization fluctuations as they can collectively hop along the backbone of the chain and they can display biased random walk. This process leads to a fluctuation in the counterion population around DNA and generates a current. As a result of the change in polarizability, the dielectric increment ($\Delta\varepsilon$) can be obtained by using Eq. (22)

$$\Delta\varepsilon = \varepsilon_0 \left(\frac{Nl}{V}\right)\left(\frac{8\pi}{3}\right) Q \sum_k \frac{1}{k^2}\left(\frac{1}{1+Q\omega_k}\right) \quad (22)$$

where $\varepsilon_0$ is the SDC of the solvent, (N/V) is the number density of the polyion chain, $l$ is the length of the chain, Q is $n\beta e^2/\varepsilon_0 l$, n is the average number of bound counterions, and $\omega_k = \frac{\varepsilon_0}{e^2}\int_0^l dr\, \varphi(r)\cos(kr)$. The contribution from the different modes approximately scales as $1/k^2$ which makes the contribution from large k negligible. The value of $\Delta\varepsilon$ varies from a few hundred to even a few thousand depending on the density.

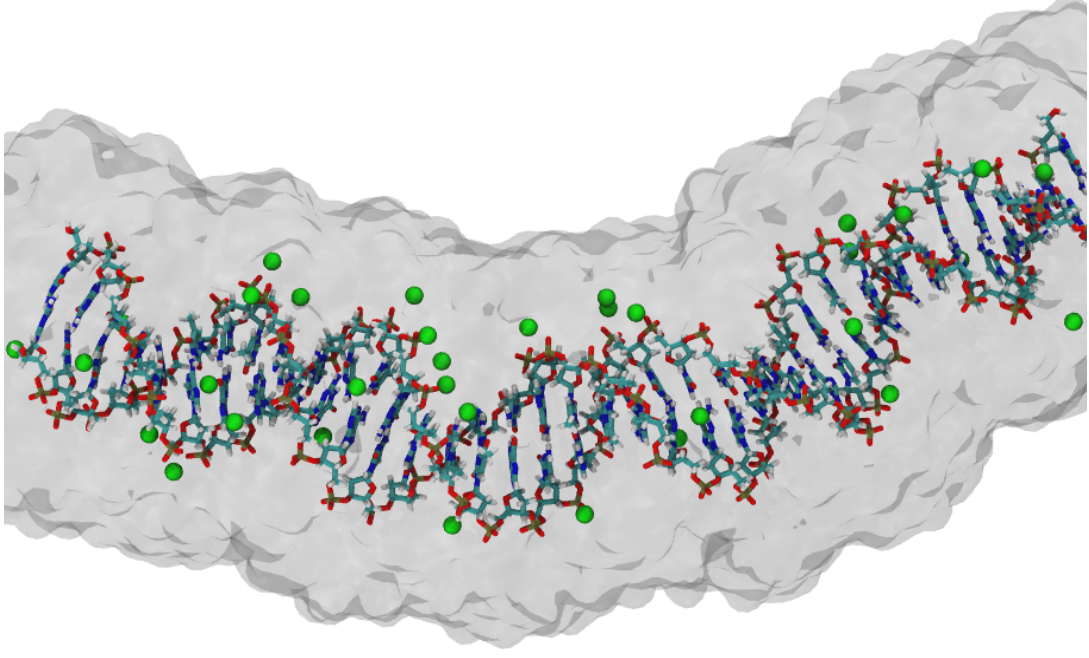

*Figure 9. A segment of double-stranded DNA with its hydration layer (transparent grey) and trapped counterions (green) on its backbone.*

The DNA-counterion system qualifies as a good example where the continuous time random walk (CTRW) theory, developed by Scher-Montroll-Lax,[72] can be applied to understand the origin of slow dynamics in aqueous DNA.[71] Because of the counterions fluctuations and the fluctuations of DNA itself, any probe attached to the DNA experiences a current [I(t)] with is proportional to the time-derivative of the solvation energy [E(t)] of the probe. In such a complex system with a spatially dispersed traps with differing energy barrier, one cannot approximate the waiting time distribution [$\psi(\tau)$] as a rapidly decaying function with a single transition rate. We rather approximated $\psi(\tau)$ as

$$\psi(\tau) \sim \exp(\tau) \int_{\sqrt{\tau}}^{\infty} dz \left(z - \sqrt{\tau}\right)^2 \exp(-z^2). \qquad (23)$$

The above description of the waiting time distribution makes the long time behaviour acquire a $\tau^{-(1+\alpha)}$ dependence, where $0 < \alpha < 1$. From here, one can follow the procedure described by Montroll and colleagues to obtain the power law dependence of I(t). It was shown that $I(t) \sim t^{-(1-\alpha)}$ when $<l> \ll L$ and $I(t) \sim t^{-(1+\alpha)}$ when $<l> \gg L$;[71] [$<l>$ and L being the mean position of the Gaussian current packet and the total chain length respectively]. As I(t) is related to the time derivative of the solvation energy [E(t)], it becomes clear that E(t) would also exhibit a power law decay (either $t^{\alpha}$ or $t^{-\alpha}$). Now, the timescale of Debye relaxation ($\tau_D$) is proportional to the solvation timescales ($\tau_L$). Therefore, the origin of slow dielectric relaxation can also be comprehended from this analysis.

## VII. Surface effects versus confinement

It is difficult to disentangle the two effects in the nanoconfinement domain. This issue was raised in the study of solvation dynamics and also two-dimensional infrared spectroscopic studies of water confined within reverse micelles.[73,74] When the confining surface is hydrophobic, as in the case of graphene surfaces employed here, one would have expected minimal effects of the role of surface-water interactions. However, that turned out to be not the case because of the occurrence of the dangling hydrogen bond of water molecules. In the reverse micelle case, water molecules face a polar or charged surface, and the nature of interactions is different. Here also one finds the occurrence of a bulk water-like pool at the centre at not too large sizes. Of course, the surface effect on the dielectric properties of the confined water have not been studied experimentally. However, it is clear that the peculiarities of water play a role because the dielectric constant of Stockmayer liquid approached bulk value at a much faster rate than water. It is not clear how one can disentangle the two effects, as both decrease as the system size is increased. One thing is certain. in the case of water, surface effects propagate inside to provide stronger confinement effects.

## VIII. Conclusions

Dielectric constant of a nanoconfined liquid is an intriguing property that is non-trivial to understand. However, certain perspectives can be developed. First, the liquid is heterogeneous on the molecular length scale. Because the static dielectric function is a two-point correlation function that in an inhomogeneous system depends on two positions $r$ and $r'$, and not just on their separation, the function can be affected strongly by the proximity of the positions of the two confining surfaces as in a narrow slit or curved surfaces inside a sphere. However, one can still define a dielectric function, as in an isotropic bulk liquid, by considering the fluctuations of the total dipole moment, which is obtained by the sum over the dipole moments of the individual molecules. However, the value of the local dielectric constant is to be defined in a self-consistent fashion where the value in one location (as in a grid) is required to be self-consistent with the rest. Second, in nanoconfined systems, the dielectric function can become anisotropic, for example, both for the nano slit and the nano tube. This total dipole moment can be used to find an effective electric polarization inside the liquid so constrained. As mentioned earlier, the nano slit can serve as a model in industrial and biological applications.

The important experiments of Geim *et al.* bring out the importance of layer-wise heterogeneity in the dielectric response of the confined liquid. Such a heterogeneity is also present in the liquid confined within a sphere and appears to be a universal property of confined dipolar liquids. We observe such heterogeneity in protein hydration layer and also in DNA grooves. It has been particularly well studied

It has been noted earlier[75,76] that single particle properties, such as the translational diffusion and rotational relaxation time constant (measured by fluorescence depolarization or NMR), approach bulk values rapidly. In fact, single particle dynamical characteristics approach those of bulk water as we move away from the surface, within 2-3 layers of water, say in the narrow slit. We already observed that in the cyclodextrin cavity, there appears a pool of water in the middle that has bulk water characteristics.[77] In sharp contrast, the collective properties display a strong system size dependence. Since the static dielectric constant (SDC) seems to be particularly sensitive to the length scale of confinement, the study of SDC seems to be particularly useful in understanding dipolar correlations in small sized systems. This sharp contrast between single particle and collective properties indicates an intriguing role of intermolecular correlations that is not yet fully understood.

Reduced value of the static dielectric constant within a sphere of the size of a few nm can have practical consequence in modelling the electrostatic interactions between two charges inside the confined water. One is reminded of the reduced dielectric constant of protein interior. Electrostatic interactions within protein are modelled by assuming a dielectric constant with values between 2 and 3. Such a small dielectric constant can have important consequences. In the case of nanoconfined spherical water, such low dielectric constant was not expected. This could be important in modelling reaction in such systems as cyclodextrin cavity. Low interfacial dielectric permittivity can partly explain the marked reaction acceleration observed in aqueous microdroplet surfaces.[78] In the case of nanoslit, the screening of electrostatic interaction becomes complex because the dielectric function is not a scalar quantity anymore, with less screening in the perpendicular direction. This anisotropy can play an important role in the relative dynamics of a pair of charges.

The basic Perspective that emerged from the works reviewed here is that the origin of the surprisingly slow (with the increase in the size of the system) approach to the bulk value of the static dielectric constant remains to be understood quantitatively. One can understand the rapid approach to the bulk of the single particle dynamical properties, like translational diffusion and rotational correlation time, in terms of the fact that these are more influenced by local environment that reverts back to the bulk at a short distance from the surface. However, the same is not true for the intermolecular correlations that determine the collective properties. They remain perturbed by the surface. We finally note that this area requires more theoretical and computational work. One interesting development would be an attempt to use the integral equations that could reveal microscopic details of two and three-particle correlations and their dependencies on surface-liquid interactions.

## Acknowledgment

BB thanks SERB (DST), India for an Indian National Science Chair Professorship. SM thanks the University Grants Commission (UGC), Indian Institute of Science (IISc), and SERB for research fellowship and research associateship, during the time of these works.